# The Instability of Markets

Tad Hogg, Bernardo A. Huberman and Michael Youssefmir

Dynamics of Computation Group
Xerox Palo Alto Research Center
Palo Alto, CA 94304

## Abstract

Recent developments in the global liberalization of equity and currency markets, coupled to advances in trading technologies, are making markets increasingly interdependent. This increased fluidity raises questions about the stability of the international financial system. In this paper, we show that as couplings between stable markets grow, the likelihood of instabilities is increased, leading to a loss of general equilibrium as the system becomes increasingly large and diverse.

June 30, 1995

## 1. Introduction

The explosive growth of computer networks and of new forms of financial products, such as derivatives, is leading to increased couplings among previously dispersed markets. This increased fluidity raises questions about the stability and efficiency of the international financial system. On the one hand, it is quite apparent that increased overall connectivity among markets allows transactions that were not previously possible, increasing the net wealth of people and leading to more efficient markets. But at the same time, this implies that a transition is taking place, from a more static scenario in which isolated markets can be considered in equilibrium on their own, to a global economy that knows of no geographical borders. One may wonder about the nature of this transition, i.e. is it gradual in the sense of a smooth change in prices, or punctuated by abrupt changes in the value of certain commodities, cascading bankruptcies and market crashes?

Underlying these questions is the old problem of the existence and stability of equilibria in markets, which general equilibrium theory has addressed under various conditions [1, 2]. In this paper, we will focus on the stability on markets by treating the couplings among them as dynamical entities in the spirit of other evolutionary approaches [6, 16, 4].

The standard approach to characterizing the stability of markets postulates that agents opportunistically optimize their portfolios in such a way so as to minimize their risk while maintaining a given return. But in real life agents are not always able to perfectly process information about the system in which they are embedded. Under such conditions, adaptive agents continuously switch between different behavioral modes in response to a constantly changing environment, favoring behaviors that can temporarily lead to increased rewards. It follows that while agents are continuously learning about the relative merits of different commodities and markets, the couplings between such commodities evolve in rather complicated ways.

Other examples of such evolving couplings are provided by agents linking limited baskets of commodities efficiently while ignoring other commodities in the process. Of more recent interest are the couplings that agents introduce through complicated derivatives created in an effort to optimize and hedge away risks. Due to their complex nature such derivatives can be poorly understood by the people who use them, once again leading to extraneous couplings. Finally, highly leveraged hedge funds can also introduce couplings into the market by being forced to cover certain leveraged positions with other unrelated positions.

In this paper, we assume the existence of an equilibrium within a network of markets and examine its dynamic stability through the use of a general model of the adjustment processes within the markets. We focus on cases where agents are not perfectly rational or identical and therefore can introduce couplings that may not always represent the best possible allocation of their resources. We then discuss how the stability of such a system scales as different markets become increasingly coupled. We show that the class of systems that are stable becomes smaller and smaller as the number of coupled markets scales up. These instabilities are shown to exist even when couplings are so weak that the markets are near decomposable. Such instabilities in turn require a heightened degree of learning on the part of market participants, a process in which the learning itself is marked by periods of instability until equilibrium is reestablished. These results, which run counter to prevailing notions of stability in large coupled markets, offer a cautionary note on treating emerging markets with the tools of equilibrium economics.

## 2. Dynamics of Coupled Markets

In what follows we will consider the case of a number of markets, each of which is in stable equilibrium when decoupled from the others. These markets contain arbitrarily large number of agents that buy and sell commodities, using a diverse set of strategies. Heterogeneous agents will provide diverse couplings between these markets through mechanisms such as physical substitutions, financial derivatives, arbitraging between geographically disperse markets, and expectations (be they rational or irrational) that movements in one market will lead to changes in others.

Given a single commodity, a reasonable dynamic model of price adjustments postulates that the price rises when demand exceeds supply and that they fall when supply exceeds demand. This can be described by a differential equation of the form

$$\frac{dp}{dt} = f(p) \tag{1}$$

where $p$ is the price of the commodity and the function $f$ represents the excess demand at the given price. This equation can be thought of as a description of market adjustments via the *tatonnement* process by which an auctioneer calls out adjustments in prices in order to satisfy excess demand. The function, $f$, then defines the equilibrium $p^0$ where $f(p^0) = 0$. The value $f'(p^0)$ represents a linearization of the adjustment process near the equilibrium. If this quantity is negative the equilibrium at $p^0$ is stable against small perturbations, which relax on a time scale given by $\tau = |f'(p^0)|^{-1}$.

The time scale $\tau$ for adjustment is reduced when transaction costs are negligible or agents are confident about the value of the commodity. Any movement of $p$ away from $p^0$ will be immediately corrected for by agents seeking to make a profit on the differential $(p - p^0)$. On the other hand, when agents are uncertain or have diversity of expectaions for the fair price for the commodity, the adjustment process may depend on agent interactions, communication, and analysis. In this case, the time scale for adjustment $\tau$ will tend to be greater than before.

These remarks are readily generalized to the adjustment processes where there is more than one commodity and multiple markets. We will study the dynamics that result from the couplings of these markets, and in particular how the couplings affect the stability of the system as whole. To do this we interpret $p$ as a price vector, whose entries correspond to prices in the separate markets, and $f$ as a vector of excess demands that characterizes both the dynamics of the isolated markets and their couplings. The dynamics is then given by an equation of the form Eq. 1, which relates the evolution of prices of all commodities in all markets as a function of each other and the degree to which they are coupled [3, 19, 15].

The excess demand in the case of each commodity is controlled by the individual agents within the respective markets. In stable equilibrium the individual excess demands of the agents are such that in totality prices are held at the equilibrium values. We can then study stability around an equilibrium $\vec{p_{eq}}$ where $\vec{f}(\vec{p_{eq}}) = 0$, by assuming the existence of a temporal departure from this equilibrium by a small amount $\vec{\delta}$. In order to see if this disturbance decays in time or grows, we perform a Taylor series expansion around the fixed point to get an equation of the form

$$\frac{d\vec{\delta}}{dt} = M(\vec{p_{eq}})\vec{\delta} \qquad (2)$$

where the $n$ by $n$ matrix, $M$, is the Jacobian of $\vec{f}$ evaluated at the fixed point $\vec{p_{eq}}$ and $n$ is the number of commodities.

The components of this Jacobian matrix describe how a small increase in the price of one item in one market changes that of another item. Thus, the diagonal elements of $M$ show the direct effect of a small change in excess demand on the price of a given commodity. We assume that each market by itself is stable so as to counteract the original change, which implies that the diagonal elements will be negative, with average value that we denote by $-D$, and which quantifies the speed of adjustments within that market.

On the other hand an off-diagonal entry describes the direct effect on an item from a change in the price of another, which can be of either sign. Notice that if the couplings between commodities are weak, it would translate into a matrix whose largest entries are on the diagonal and the off-diagonal terms would be small. As discussed above, these off-diagonal couplings are the result of individual agents whose expectations, whether informed or uninformed, link the two commodities such that an increase in one commodity's price affects that of the other.

The effect of slight disturbances around the fixed point perturbation is determined by the eigenvalue of $M$ with the largest real part, which we denote by $E$. Specifically, the long time behavior of the perturbation is given by $\delta \propto e^{Et}$, implying that if $E$ is negative, the disturbance will die away and the system will return to its original equilibrium. If, on the other hand, $E>0$, the smallest perturbation will grow rapidly in time, leading to instability.

We note that this model of dynamic adjustment and its stability has also been used in the economic literature focusing on qualitatively specified matrices [17]. In that literature, the focus is on specifying the stability properties of matrices given only the qualitative nature of the signs of the entries (positive, negative, or zero). In that case, for example, commodities that are gross substitutes would be coupled via positive nondiagonal matrix elements while gross compliments would be coupled by negative coefficients. Such an approach, however, suffers from the disadvantage that the cases for which stability criterion are specified are quite restricted.

## 3. Market Instabilities

The approach taken in this paper looks at the average behavior of market given an ensemble specifying uncertain knowledge of parameters within the stability matrix. For the sake of treating a very general case, we will assume as little knowledge about these mechanisms as it is possible. This implies that all matrices that are possible Jacobians can be considered, and that there is no particular basis for choosing one over the other. This is the class of the so-called random matrices, in which all such matrices are equally likely to occur. Matrices in this class are such that each entry is obtained from a random distribution with a specified mean and variance. By taking this approach, we can make general statements about the stability of these systems as the number of markets and diversity of agent behaviors grows. We show quite generally that as the number of coupled markets and the diversity among the agents grows, it is more and more likely that the system as a whole will be unstable.

The precise value of the largest eigenvalue $E$ depends on the particular choice of the Jacobian matrix. Methodologically, the study of the general behavior of these matrices is performed by examining the average properties of the class that satisfies all the known information about them. A class of plausible stability matrices is determined by the amount of information one has about particular market mechanisms and their couplings. This information, which is far from perfect, depends on the nature of expectations that agents have about future values, which are based on limited knowledge, and on the extent to which individuals understand the nature of the financial instruments that they invest in.

In spite of their random nature, these matrices possess a number of well defined properties, among them the behavior of their eigenvalues [22, 14, 5, 8, 11, 7]. This means that we can use these properties to ascertain the stability of the markets against perturbations in the excess demand. In what follows we will show that in the general case, as these matrices grow in size, or the variability of their entries, their largest eigenvalue becomes positive, thus leading to market instability. This is a result that applies not only to markets but also to complex ecosystems [12, 13, 9].

The simplest case, albeit not very realistic, would correspond to a situation where commodities are equally likely to be substitutes or complements of each other. This means that on average the nondiagonal elements would be of zero mean. We will also assume that the Jacobian random matrix will have symmetric entries and bounded in magnitude. For this case, the distribution of eigenvalues as a function of the size of the Jacobian is given by Wigner's law [22], i.e.

$$E = 2\sigma\sqrt{n} - D \tag{3}$$

This implies that as the matrix gets large enough, its largest eigenvalue will become positive.

A more realistic case relaxes the requirement that the changes in one price are on average balanced by changes in others. This corresponds to having a non-zero mean. In this situation a theorem of Furedi and Komlos [8] states that, on average, the largest eigenvalue is given by

$$E = (n-1)\mu + \sigma^2/\mu - D \tag{4}$$

where $\mu$ is the average value of the couplings, which we assume to be positive. Moreover, the actual values (as opposed to the average) of $E$ are normally distributed around this value with variance $2\sigma^2$. This implies, that as the size of the system grows, most such matrices will have positive largest eigenvalue, thus making the equilibrium point unstable.

Consider next the case of a non-symmetric matrix, with nondiagonal terms with positive mean, $\mu$, standard deviation, $\sigma$, and whose diagonal terms have mean, $-D$. In this case the largest eigenvalue grows with the size of the matrix as [8, 11]

$$E \sim \mu(n-1) - D \tag{5}$$

Since $\mu$ is positive these results imply that even if markets are stable when small, they will become unstable as their size becomes large enough for $E$ to change sign.

One argument that could be given for the stability of coupled markets in spite of their size is that not all commodities happen to be coupled to each other. In terms of our theory, this amounts to a *near decomposability* of the Jacobian matrix whereby the entries are such that the further they are from the diagonal the smaller they become [20]. Such systems, sometimes called loosely coupled, are very relevant to situations when markets are initially weakly coupled to each other. But as we will now show, even in this case, as the size of the markets increases the equilibrium will be rendered unstable. In terms of the interactions, this situation can describe either the fact that a given item's price is strongly influenced by a few others and weakly by the rest, or a more structured clustering, where the commodities appear in groups (e.g. technology stocks, foreign currencies) wherein their members strongly interact but members of different groups have weak interactions with each other.

Another way of considering this scenario is to imagine initially isolated markets that eventually get coupled through the interaction of mediating interactions, such as roads and communications. In this case the coupling between the initially isolated markets grows in time.

The corresponding matrices for the first case are constructed by selecting off-diagonal entries at random with large magnitude 1 with probability $p$ and small value $\epsilon < 1$ otherwise. In this case $\mu = p + (1-p)\epsilon$ and $\sigma^2 = p(1-p)(1-\epsilon)^2$. As shown in Eq. 5, the largest eigenvalue will become positive when the system becomes large enough.

In the second case, the commodities are grouped into a hierarchical structure which we assume to be of depth $d$ and average branching ratio, $b$. In this representation, the strength of the interaction between two commodities will decrease based on the number of levels in the hierarchy that one has to climb to reach a large enough common group to which they both belong. Specifically, the interaction strength will be taken to be $R^h$, with $h$ the number of hierarchy levels that separate the two commodities, and $R$ characterizes the reduction in interaction strength that two commodities undergo when they are separated by one further level. The average size of the matrix is given by $n = b^d$ and the mean of the non-diagonal terms can be computed to be

$$\mu = \frac{\sum_{h=1}^{d} b^{h-1}(b-1)R^h}{\sum_{h=1}^{d} b^{h-1}(b-1)} = \frac{(Rb)^d - 1}{b^d - 1} \frac{R(b-1)}{Rb - 1} \tag{6}$$

In order to represent a reasonable clustering, we need to specify whether the total interaction is dominated by either those neighboring commodities in the hierarchy we choose, or distant ones. In the first case, this amount to restricting $R$ to be $R < 1/b$. Notice that this choice makes the decreasing interaction strength between commodities overwhelm the increase in their number as higher levels in the hierarchy are considered.

In this situation when $n$ is large Eq. 6 becomes $\mu = \frac{R(b-1)}{(1-Rb)n}$, which implies that as $n$ grows the mean goes to zero, leading to a stable system because of Eq.5. Notice however, that if fluctuations in the coupling strength were taken into account, some of the matrices could still become unstable, as was shown above.

In the second case, $R > 1/b$. This implies that in the large $n$ limit the average becomes

$$\mu = R^d \frac{R(b-1)}{Rb-1} = \frac{R(b-1)}{Rb-1} n^{\frac{\ln R}{\ln b}} \tag{7}$$

Note that since $-1 < \frac{\ln R}{\ln b} < 0$, $\mu$ goes to zero as the size of the matrix grows, but in slower fashion than the case above. Nevertheless, this subtle difference in convergence to zero amounts to a qualitative difference in the stability of the system. To see this, notice that Eq. 5 implies that the largest eigenvalue of a random matrix grows as $\mu n$, which increases with $n$ for this value of $\mu$, thus leading to instability when the system gets large enough. The growth in largest eigenvalue with the size of the system is exhibited in figure 1 for a particular choice of parameters. Notice that the system becomes unstable for $d \geq 5$. Given these results we see that the size of the matrix for which this instability takes place is much larger than the one in the absence of a hierarchy of interactions.

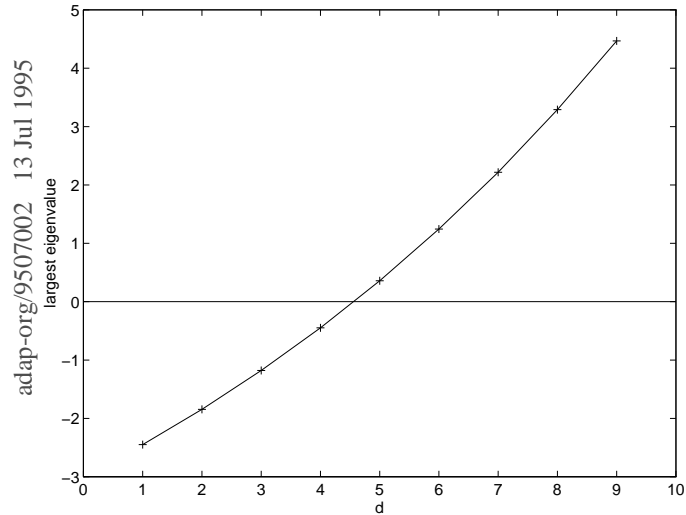

**Fig. 1.** Plot showing the growth of the largest eigenvalue of a hierarchical matrix with branching ratio $b = 2$ and $R = 0.55$ as a function of $d = \log_2 n$. The diagonal elements were chosen to be $D = 3$. The points are the theoretically predicted values of Eq.6 and lie very close to the computed eigenvalues shown by the solid line.

A final possibility is for the commodities to include aggregate structures, such as stock indices. In this case the couplings will correspond to situations where each commodity interacts with itself, the other components of its aggregate, or its higher order aggregate. The ensuing Jacobain will have blocks of nonzero elements and zero entries elsewhere. For this case, we have shown that an even slower growth of the largest eigenvalue with size is obtained [9]. Specifically, the eigenvalues grow no faster than $\sqrt{\ln n}$, implying that much larger coupled markets can be stable when they are structured in a highly clustered fashion. Fluctuation corrections make the eigenvalues grow like $\ln n$, still implying a higher degree of stability than in the previous cases.[1].

## 4. Discussion

Recent dramatic fluctuations and losses in the world financial markets have raised concerns about the inherent stability of the global financial system. These concerns have been prompted by the heightened fluidity of global currency flows and the emergence of complicated derivatives, which now allow market players to make financial bets as never before. For example, the derivatives debacle in Orange County, California points to the fact that some market players do not understand the full risks that are being taken. At the same time, the result of the current trends in global finance is that markets are now more and more coupled and individual governments have less and less power to control these perhaps destabilizing couplings. The recent economic crisis in Mexico, the "tequila effect", resulted in added volatility and "corrections" in many emerging equity markets all over the world.

---

[1] A further interesting possibility [21] is that the linearized system can produce an initially growing transient even when $E < 0$ so that eventually the perturbations decay. During this transient growth, the perturbations may become large enough to be sustained by nonlinear corrections, thus giving another possible source of instability.

In this paper, we modeled this system as a web of interacting markets much like a biological ecosystem [18]. By doing so we obtained a general result showing that as couplings between previously stable markets grow, the likelihood of instabilities is increased.

These results allow us to understand phenomena that are likely to arise as a system grows in diversity, strength of couplings and in the size of the overall number of the coupled markets. In a sense these results appear to be counterintuitive, for one expects that as a system gets larger, disturbances in a particular part would exhibit a kind of decoherence as they propagate through the system, making it very unlikely that after a given time they would once again concentrate on a particular node and amplify it. But the properties of random matrices make it probable for this conspiracy of perturbations to concentrate on given parts of the market. As the size and diversity of couplings in such matrices grow, the complicated effects of the couplings are more and more likely to result in instability that leads to motion away from equilibrium.

One may ask about the fate of the lost equilibrium and the ensuing evolution of an unstable market. We speculate that once the instability sets in adaptive agents will modify the respective couplings in such a way as to stabilize the entire system once more at another equilibrium. If this is indeed the case, the volatility brought about by instabilities in such large systems is the natural mode by which couplings are modified to achieve a more efficient market. Due to the lack of information needed for appropriate centralized control, it is also by no means clear that regulatory approaches to controlling the market structures will be effective. Improper controls could introduce additional couplings in such an uncertain way that the system may be further destabilized. Accepting the instabilities of these larger systems as the best way for market participants to learn the correct couplings may, therefore, be the most reasonable course of action.

Last but not least, these results cast light on the related problem of the dynamics of multiagent systems in distributed computer networks, which have been shown to behave like economic systems [10]. In the case of only two resources their equilibrium is punctuated by bursts of clustered volatility [23], a fact which renders the notion of equilibrium suspect. This paper shows that as distributed computing systems get large and more coupled, they will also exhibit a loss of stability, on their way to finding a new and more efficient equilibrium.

# References

[1] K. J. Arrow and F. H. Hahn. *General Competitive Analysis*. Holden-Day, San Francisco, 1971.

[2] Kenneth J. Arrow. Workshop on the economy as an evolving complex system: Summary. In P. W. Anderson, K. J. Arrow, and D. Pines, editors, *The Economy as an Evolving Complex System*, pages 275–281. Addison-Wesley, 1988.

[3] Kennneth J. Arrow and Leonid Hurwicz. On the stability of competitive equilibrium. *Econometrica*, 26:522–552, 1958.

[4] W. A. Brock. Nonlinearity and complex dynamics and economics and finance. In P. W. Anderson, K. J. Arrow, and D. Pines, editors, *The Economy as an Evolving Complex System*, pages 77–97. Addison-Wesley, 1988.

[5] J. E. Cohen and C. M. Newman. The stability of large random matrices and their products. *Annals of Probability*, 12:283–310, 1984.

[6] Richard H. Day. Adaptive processes and economic theory. In Richard H. Day and Theodore Groves, editors, *Symposium on Adaptive Economics*, pages 1–35. Academic Press, 1975.

[7] Alan Edelman. *Eigenvalues and Condition Numbers of Random Matrices*. PhD thesis, MIT, Cambridge, MA 02139, May 1989.

[8] Z. Furedi and K. Komlos. The eigenvalues of random symmetric matrices. *Combinatorica*, 1:233–241, 1981.

[9] T. Hogg, B. A. Huberman, and Jacqueline M. McGlade. The stability of ecosystems. *Proc. of the Royal Society of London*, B237:43–51, 1989.

[10] Bernardo A. Huberman and Tad Hogg. Distributed computation as an economic system. *J. of Economic Perspectives*, 9(1):141–152, 1995.

[11] F. Juhasz. On the asymptotic behavior of the spectra of non-symmetric random (0,1) matrices. *Discrete Mathematics*, 41:161–165, 1982.

[12] R. M. May. Will a large complex system be stable? *Nature*, 238:413–414, 1972.

[13] Ross E. McMurtrie. Determinants of stability of large randomly connected systems. *J. Theor. Biol.*, 50:1–11, 1975.

[14] M. L. Mehta. *Random Matrices and the Statistical Theory of Energy Levels*. Academic Press, New York, 1967.

[15] Lloyd A. Metzler. Stability of multiple markets: The Hicks condition. *Econometrica*, 13(4):277–292, 1945.

[16] Richard A. Nelson and Sydney G. Winter. *An Evolutionary Theory of Economic Change*. Harvard University Press, 1982.

[17] James Quirk. Qualitative stability of matrices and economic theory: A survey article. In H. J. Greenberg and J. S. Maybee, editors, *Computer-Assisted Analysis and Model Simulation*, pages 113–164. Academic Press, NY, 1981.

[18] Michael L. Rothschild. *Bionomics: economy as ecosystem*. Henry Holt and Company, 1990.

[19] Paul A. Samuelson. The stability of equilibrium:comparative statics and dynamics. *Econometrica*, 9:111, 1941.

[20] Herbert A. Simon and Albert Ando. Aggregation of variables in dynamic systems. *Econometrica*, 29(2):111–138, 1961.

[21] Lloyd N. Trefethen, Anne E. Trefethen, Satish C. Reddy, and Tobin A. Driscoll. Hydrodynamic stability without eigenvalues. *Science*, 261:578–584, July 30 1993.

[22] E. P. Wigner. On the distribution of the roots of certain symmetric matrices. *Annls Math*, 67:325–327, 1958.

[23] Michael Youssefmir and Bernardo A. Huberman. Clustered volatility in multiagent dyanmics. *Journal of Economic Behavior and Organizations*. to appear.